\documentclass[prd, amsfonts, twocolumn, nofootinbib, showpacs]{revtex4}
\pdfoutput=1
\usepackage[plainpages=false, colorlinks=true, anchorcolor=blue, linkcolor=blue, citecolor=blue, bookmarks=false]{hyperref}
\usepackage{graphicx}
\usepackage{epsfig}
\usepackage{color}
\usepackage{amsmath}
\usepackage{amssymb}
\newcommand{\dg}{$\frac{\delta g}{\bar{g}}$ }
\newcommand{\be}{\begin{equation}}
\newcommand{\ee}{\end{equation}}
\newcommand{\bea}{\begin{eqnarray}}
\newcommand{\eea}{\end{eqnarray}}

\newcommand{\gapp}{\mathrel{\raise.3ex\hbox{$>$}\mkern-14mu
\lower0.6ex\hbox{$\sim$}}}
\newcommand{\lapp}{\mathrel{\raise.3ex\hbox{$<$}\mkern-14mu
\lower0.6ex\hbox{$\sim$}}}
\def\bbox{{\,\lower0.9pt\vbox{\hrule \hbox{\vrule height 0.2 cm
\hskip 0.2 cm \vrule  height 0.2 cm}\hrule}\,}}
\newcommand{\rthis}[1]{\textcolor{black}{#1}}

\begin{document}
\title{Search for variability in Newton's constant using local gravitational acceleration measurements}
%\author{}
%\affiliation{ }
%\author{Srinikitha Bhagvati$^a$\ Shantanu Desai$^{a,1}$

\author{Srinikitha \surname{Bhagvati} }%
\altaffiliation{ep18btech11003@iith.ac.in}
\author{Shantanu \surname{Desai}}
\altaffiliation{shntn05@gmail.com}

\affiliation{Department of Physics, Indian Institute of Technology, Hyderabad, Telangana-502284, India}

 %%%%%%%%%%%%%%%%%%%%%%%%%%%%%%%%%%%%%%%%%%%%%%%%%%%%%%%

\begin{abstract}

In a recent work, ~\citet{Dai} searched for a variability  in Newton's constant $G$ using the IGETS based gravitational acceleration measurements.
However, this analysis, obtained from $\chi^2$ minimization,  did not incorporate
the errors in the gravitational acceleration measurements. We carry out a similar search with one major improvement, wherein  we incorporate these aforementioned errors.
To model  any possible variation in the gravitational acceleration, we fit the  data to
four models: a constant value, two sinusoidal models, and finally, a linear model for the variation of gravitational acceleration.  We find that none of the four models provides a good fit to the data, showing that there is no evidence for a periodicity or a linear temporal variation in the acceleration measurements.  We then redid these analyses after accounting for an unknown intrinsic scatter. After this, we find that although a constant model is still favored over the sinusoidal models, the linear variation for $G$ is marginally preferred over a constant value, \rthis{using information theory-based methods}.

\end{abstract}

%%%%%%%%%%%%%%%%%%%%%%%%%%%%%%%%%%%%%%%%%%%%%%%%%%

\maketitle

\section{Introduction}
It is known for a while that the different measurements of Newton's Gravitational constant ($G$) don't agree with each other~\cite{Bethapudi,Mohr,Gillies}. There has also been a claim for a 5.9 year periodicity in $G$~\cite{Anderson}, although this was rebutted using both Bayesian and frequentist analysis~\cite{Pitkin,Desai16} (see also ~\cite{Newman}).  Some precision gravity experiments have also found tentative deviations from Newtonian gravity~\cite{Krishak}, which could be connected to some of the unsolved problems in Cosmology such as Dark Energy, Dark Matter, Hubble tension, etc~\cite{Periv}. \rthis{ However, we note that a large number of  searches for  the variation of $G$ have also been carried out using a plethora of astrophysical observations, such as helioseismology, lunar laser ranging, Type 1a Supernovae, binary pulsars, all of which have obtained null results~\cite{Will06,PDG}. From these measurements, the variance in $\frac{\dot{G}}{G}$ is less than $10^{-12}$~\cite{Will06}. At face value, these limits already rule out any putative variations in $G$, such as those claimed in ~\cite{Anderson}. However, any limit on the variation of $G$ from astrophysical or cosmological observation always has some uncertainty and model-dependence  in it. Therefore,  it  behooves us  to search for any variation or periodicity  in $G$
using independent terrestrial probes.}

Motivated by these considerations, ~\citet{Dai} (D21, hereafter)  reconsidered the variability of $G$ using the variance of local 
Newtonian gravitational acceleration measurements ($\delta g/\bar{g}$). The key premise behind this work, was that the variance of  $G$ can be bounded from the gravitational acceleration measurements:
$$\frac{\delta G}{\bar{G}}\lessapprox\frac{\delta g}{\bar{g}}$$
where $$\frac{\delta g}{\bar{g}} = \frac{g-\bar{g}}{\bar{g}}$$ Here, $\bar{g}$ refers to the mean of the  gravitational acceleration during the observational period.

D21 used the \dg  data  compiled by the International Geodynamics and Earth Tide Service (IGETS), using data from 16 observational stations.
They considered two models for the variation of $G$, namely a sinusoidal and a linear model for the variation of the relative acceleration. D21 constructed a $\chi^2$ functional based on the residuals between  the observed data and various theoretical models. For the sinusoidal model, they looked for a minima at 5.9 years. For the linear model, the value of the slope was constrained from the likelihood distribution.  Thereafter, the relative amplitude of the sinusoidal model was constrained to be less than $3 \times 10^{-9}$, whereas for the linear model $\dot{G}/G < 10^{-10} \rm{yr^{-1}}$. 
In D21, the observed errors in \dg  were not considered for the analysis. The  reason for this is that the  variances of \dg  contain contributions not  only from the 
acceleration measurement equipment, but also from local environment. These measurement errors could  be smaller than the fluctuations from the local environmental effects. Therefore, since D21 were looking for long term trends in the variation of $G$, the measurement errors were ignored.

However, the median and mean fractional errors in \dg were about 6\% and 64\% respectively with 38 data points having fractional errors  greater than 100\%.
Furthermore, ignoring these errors we  cannot get an independent assessment of the goodness of fit   or robust  error estimates on the model parameters. 
Therefore, it is imperative to redo the analysis in D21,  after incorporating the measurement errors in the analysis. This also allows us independently test if the total errors are under-estimated or not.

In this work, we therefore revisit this issue by analyzing the same data (kindly provided to us by Dr. Dai), and 
by incorporating the errors in \dg, while constructing the  likelihood. We test for periodicities using two sinusoidal functions as well as a  linear model for the variation in \dg. \rthis{We note, however, that the IGETS measurements analyzed in this work are not the only \dg measurements available. There also exist gravitational acceleration measurements from several hundred FG5 gravimeters distributed around the world. We shall also analyze these data in a future work.}

\rthis{The manuscript is organized as follows.}
Our analyses  is described in Sect.~\ref{sec:analysis}  and Sect.~\ref{sec:scatter}, wherein we account for an additional unknown systematic error. We conclude in Sect.~\ref{sec:conclusions}. \rthis{We have also made our analysis codes used in this work publicly available.}

\section{Analysis}
\label{sec:analysis}
\begin{figure}
        \centering
        \includegraphics[scale=0.6]{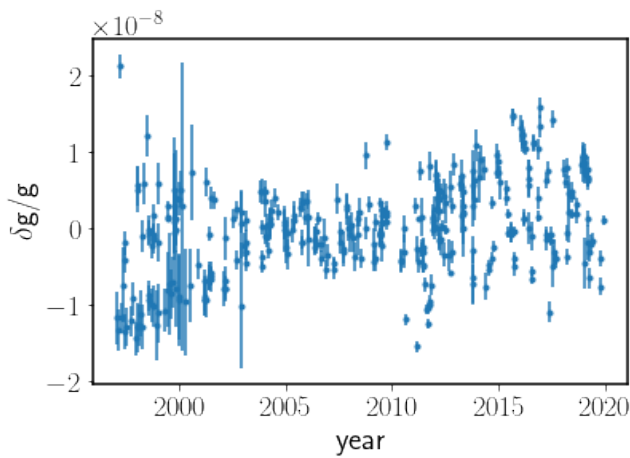}
        \caption{Time-series measurements of relative gravitational acceleration based on IGETS measurements collated by Dr. Dai and used in D21.}
        \label{fig:delta_g_vs_year}
\end{figure}

The data showing the \dg measurements, which we use for our fits is the same as in D21, and is shown in Fig.~\ref{fig:delta_g_vs_year}. We  fit the data to the following four functions. The first function is a constant model. We then consider two sinusoidal functions, with the first one consisting of a mixed sine+cosine, and the second one consisting of  sinusoidal with a phase. The last model we consider is a linear  model. We enumerate  these four functions below:
\begin{eqnarray}
y(t) &=& C  \label{eq:const}\\
y(t) &=& a+A_c \cos(\omega t) + A_s \sin (\omega t) \label{eq:sinu1}\\
y(t) &=& a+A \cos(\omega t + \phi) \label{eq:sinu2} \\
y(t) &=& a+ b (t-2000) \label{eq:linear}
\end{eqnarray}

Note that Eq.~\ref{eq:sinu1} and Eq.~\ref{eq:linear} have also been used by D21. Eq.~\ref{eq:sinu2} is a variant of the sinuosoidal model used in D21, with a single sinusoid along with a phase. A similar model is usually  used to fit annual modulation in dark matter experiments~\cite{Dantuluri,Srinikitha}.
We find the best-fit parameters for each of these models using the following likelihood ($L$)
\begin{equation}
-2\ln L = \large{\sum_i} \ln 2\pi\sigma_i^2 + \large{\sum_i} \frac{[y_i-y(t)]^2}{\sigma_i^2},
\label{eq:lnL}  
\end{equation}
where $y_i$, $\sigma_i$ denote the \dg measurements and their errors, respectively; $y(t)$ represents the model proposed for the variation of \dg.
For all the models (except the constant model) the sampling of this likelihood has been done using the {\tt emcee}~\cite{emcee} MCMC sampler.

For each of the models, we construct  the allowed regions in case of multiple parameters and tabulate the  best-fit $\chi^2$ in Table~\ref{tab:my_label}. We now present our results for each of the models.

\begin{enumerate}
    \item \textbf{Constant model}
    For a constant model, the maximum likelihood solution is the weighted mean. The best-fit constant value is equal to  $(-9.89 \pm 6.2 ) \times 10^{-11}$. 
    The minimum $\chi^2$ value is  shown in Table~\ref{tab:my_label}. As we can see that the $\chi^2$/dof is greater than 20,  implying that  the constant model is  not a good fit to the data.
    
    \item \textbf{First sinusoidal model - Eq.~\ref{eq:sinu1}}
    We now do a fit to Equation~\ref{eq:sinu1}. This function is almost the same as that used in D21. The main difference being that in D21, a separate $a_i$ was used for every station, whereas here we assume $a_i$ is the same between all the stations. However, one caveat because of this assumption is that the uncertainty in $\bar{g}$, which is estimated for every station introduces an unknown shift in $a_i$.
     Since this model contains a sinusoidal function, the lower bound on the period should be greater than the sampling time, in order to avoid the overfitting problem which may happen for sinusoidal functions~\cite{Melchior,Krishakcosine}. The median sampling time is 12.4 days, while the mean sampling time is 23.5 days. Therefore, we choose uniform priors on $\omega$, viz. $\omega \in [0.55 , 45.92]$ rad/yr corresponding to a period of 11.4 years (half the total duration of the  dataset) and 50 days, respectively. This range is broad enough to cover the previously observed peak at 5.9 years.  
    We also choose uniform priors on $a$, $A_c$, and $A_s$  given by:  $a \in [-2 \times 10^{-8}, 2 \times 10^{-8}]$, $A_c \in [0, 2 \times 10^{-8}]$, and   $A_s \in [0, 2 \times 10^{-8}] $. 
    
   Our results showing the 68\%, 90\% and 95\% credible intervals for the free parameters of this model are shown in Fig.~\ref{fig:contour_sinu1}. We find that the marginalized credible intervals for $\omega$ show four peaks. These values are tabulated in Table \ref{tab:my_label}. The best-fit values for the other parameters for the value of $\omega$ with the maximum likelihood  are $a = (-2.97^{+1.00}_{-2.49}) \times 10^{-10}$, $A_c  = (0.12^{+0.10}_{-1.98}) \times 10^{-9}$, and $A_s = (2.14^{+2.03}_{-0.15}) \times 10^{-9}$.
None of the peaks in $\omega$ correspond to a period of 5.9 years. The $\chi^2$ per degree of freedom for all these peaks are again greater than 20, indicating that none of the peaks in $\omega$ provide a good fit to the data.
   
Similar to D21, we find that $\sqrt{A_c^2+A_s^2}$ for the best-fit likelihood is equal to $\sim 2.1 \times 10^{-9}$,  and hence  is about four orders of magnitude more stringent than the periodicity ascertained from previous $G$ measurements~\cite{Anderson}.
   
\item  \textbf{Second sinusoidal model - Eq.~\ref{eq:sinu2}}
We now do a similar fit to the second sinusoidal model listed  in Eq.~\ref{eq:sinu2}, which involves a phase factor. The priors for $a$, $w$, and $A$ are the same as those for $a$, $w$, and $A_c$, respectively used in Eq.~\ref{eq:sinu1}. For $\phi$, we use a uniform prior given by $\phi \in [0, 2\pi]$ . The best-fit values at the maximum likelihood  obtained for these parameters are: $a = (-3.12^{+0.84}_{-2.32}) \times 10^{-10}$, $A = (2.24^{+0.25}_{-0.12}) \times 10^{-9}$, and $\phi  = (4.30 \pm 0.5)$ rad.
The credible regions for the parameters of  this model can be found in Fig.~\ref{fig:contour_sinu2}. Once again, we find four peaks for $\omega$, whose values are listed in Table~\ref{tab:my_label}. Again, we don't see a peak at $\omega$, corresponding to the period of  5.9 years. Furthermore, the $\chi^2$ per degree of freedom for all the $\omega$ peaks are again greater than 20, indicating that none of the $\omega$ peaks provide an adequate fit to the data. The best-fit value of $A$ is $\sim 2.2 \times 10^{-9}$. Therefore, for both the sinusoidal models analyzed, we get a limit, which is  about four orders of magnitude more stringent than the amplitude of the sinusoidal variation inferred from   $G$ measurements.

\item \textbf{Linear Model - Eq.~\ref{eq:linear}}
Finally, we fit the data to the linear model, specified in Eq.~\ref{eq:linear}, which was also used in D21. The priors which we use for $a$ and $b$ are $a \in [-2 \times 10^{-8}, 2 \times 10^{-8}]$ and $b \in [-2 \times 10^{-8}, 2 \times 10^{-8}]$ /year, respectively.
The best-fit credible regions are shown in Fig.~\ref{fig:contour_linear}. The best fit values for $a$ and $b$ are: $a = (-0.23 \pm 0.01) \times 10^{-8}$ and $b = (2.03 \pm 0.11 ) \times 10^{-10}$ /year.  The best-fit $\chi^2$ values can be found in Table~\ref{tab:my_label}. Once again, we find that $\chi^2$/dof  values are greater than 20, indicating that the linear model is also not an adequate fit. \rthis{The limits which we obtain are not as stringent as those using astrophysical observations, although ours have been obtained using only terrestrial measurements.}

Then, using  the same {\it ansatz} as D21,   we can constrain the linear variation in $G$ using: $$\frac{\dot{G}}{G} \lessapprox b+\sigma_b = 2.14 \times 10^{-10} \text{yr.} $$ This is about twice more stringent than the limit obtained in D21.

\end{enumerate}

Therefore, in summary we find that none of the three time-dependent models (two sinusoidal based functions and a linear model) provide a good fit to the data, once we incorporate the errors in the measurements. However, even the constant model is not a good fit. The only way to get a good fit would be to add an additional systematic uncertainty (or intrinsic scatter) to the observed errors so that the $\chi^2$/dof is close to 1, similar to the analysis done  in ~\cite{Desai16,Pitkin}.

\section{Analysis with intrinsic scatter}
\label{sec:scatter}
Since the $\chi^2$/dof for all the four models considered is much greater than one, one possibility is that   the measurement errors are underestimated and there are additional sources of systematic errors. Some examples of these are enumerated in D21, and include tides, polar motion of the Earth, instrumental effects related to the speed of light, gradient height and vertical transfer, etc. Other possibilities include environment effects such as post-glacial rebound~\cite{Bilker} or water table changes~\cite{Kazama}. A detailed modelling of these additional systematics is  beyond the scope of this work.
Therefore, we redo our search for variability by adding an unknown intrinsic scatter term ($\sigma_{scatter}$) in quadrature to the measurement errors and marginalize over it, similar to  various cosmological analyses~\cite{Pradyumna}. The premise is that this intrinsic scatter encapsulates all other unknown systematic errors in the dataset.
We use log-uniform priors for $\sigma_{scatter} \in [10^{-9}, 10^{1}]$. We also carry out model comparison between the constant and the time-varying models using AIC and BIC information theory-based tests~\cite{Krishak}, where AIC/BIC are defined as follows:
\begin{eqnarray}
AIC & = & \chi^2 + 2k\\
BIC & = & \chi^2 + k\ln (n)
\end{eqnarray}
where $k$ and $n$ correspond to  the number of free parameters and the number of data points, respectively. The models with smaller AIC/BIC values are the favored ones and the significance can be assessed using qualitative strength of evidence rules~\cite{Krishak}.

The $\Delta$AIC and the $\Delta$BIC values, calculated with respect to the constant model, along with the $\chi^2/DOF$ values are given in Table \ref{tab:scatter_res}. We now summarize our results for each of these models with the addition of intrinsic scatter.

\begin{enumerate}
    \item \textbf{Constant Model (with intrinsic scatter)}:
     We choose uniform priors for $C \in [-2 \times 10^{-8}, 2 \times 10^{-8}]$, where $C$ is the constant value for \dg.  The best-fit parameters are $C = (1.27 \pm 3.20) \times 10^{-10}$ and $\log_{10}(\sigma_{scatter}) = (-8.24 \pm 0.02)$. The best-fit $\chi^2/DOF$ values are close to 1.
    
    \item \textbf{First Sinusoidal Model (with intrinsic scatter)}:
    For this analysis also, we  includ negative priors on $A_c$ and $A_s$. Uniform priors are taken for $A_c \in [-2 \times 10^{-8}, 2 \times 10^{-8}]$, and   $A_s \in [-2 \times 10^{-8}, 2 \times 10^{-8}] $. The priors on $a$ and $\omega$ are the same as those mentioned in Section \ref{sec:analysis}.  The best-fit values are $a = (1.76^{+3.26}_{-3.10}) \times 10^{-10}$, $A_c = (-0.59^{+0.97}_{-1.21}) \times 10^{-9}$, $A_s = (-1.32^{+1.23}_{-1.82}) \times 10^{-10}$, and $\log_{10}(\sigma_{scatter}) = (-8.25 \pm 0.02)$. 
    We see one prominent peak for $\omega$ occurring at $\omega = 6.10$ rad/year, and another small peak occurring at $\omega = 25.59$ rad/year, if we ignore the peak occurring close to the lower prior on $\omega$, corresponding to a time-period of $11.4$ years. This is contrary to the multiple peaks observed in Figure \ref{fig:contour_sinu1}, whose biggest peak occurs at $\omega = 24.47$ rad/year. Once again, we do not find any peak corresponding to the period of 5.9 years.

    \item \textbf{Second Sinusoidal Model (with intrinsic scatter)}:
    The prior for $A$ is the same as the prior used on $A_c$, and the priors on $a$, $\omega$, and $\phi$ remain the same as those considered for the analysis in Sect~\ref{sec:analysis}.  We find  a peak in $\omega$  at $\omega = 6.11$ rad/year, if we ignore the peak corresponding to the $T = 11.4$ years (close to the lower bound on $\omega$). This is different from the two peaks observed in the analysis without intrinsic scatter where the biggest peak occurs at $\omega = 24.50$ rad/year, as shown in Figure \ref{fig:contour_sinu2}.
    Again, we do not find a peak in the likelihood at 5.9 years.
    The best-fit values of the other parameters are $a = (1.91 \pm 3.06) \times 10^{-10}$, $A = (0.16^{+2.17}_{-2.38}) \times 10^{-9}$, $\phi = (3.38^{+1.95}_{-1.72})$, and $\log_{10}(\sigma_{scatter}) = (-8.25 \pm 0.02)$.

    \item \textbf{Linear Model (with intrinsic scatter)}:
    The priors on $a$ and $b$ remain the same as those taken previously in Section \ref{sec:analysis}. The best-fit values are $a = (-3.28 \pm 0.51)\times 10^{-9}$, $b = (3.67 \pm 0.45)\times 10^{-10}$, and $\log_{10}(\sigma_{scatter}) = (-8.28 \pm 0.02)$.
\end{enumerate}

Therefore, the best-fit intrinsic scatter which we obtain for all the four models is $10^{-8}$ and is about 2.8 times the average error in \dg measurements. With the addition of intrinsic scatter, the best-fit $\chi^2$/dof is now close to one. However, this is a consequence of fitting for an intrinsic scatter in the likelihood.
If we compare the constant model to the two sinusoidal models, we find that the constant model is decisively favored over the sinusoidal model (with $\Delta$ AIC/BIC $> 10$). However, when we compare the constant model with the linear model, we find that the linear time-varying model is marginally favored over the constant model. Since both $\Delta$AIC and   $\Delta$BIC values between the constant and linear model are less than 10, additional data would be necessary to ascertain if this difference persists and  is significant.

\section{Conclusions}
\label{sec:conclusions}
In a recent work, D21 analyzed the IGETS \dg measurements, in order to assess the  variability and periodicity in $G$ measurements previously  reported in literature~\cite{Anderson}. \rthis{Another impetus for searching for a variation in $G$ has been to address cosmological questions such as Dark energy, Dark matter, Hubble tension, etc.
Although, the limits on variation of $G$ from  astrophysical observations~\cite{Will06} already rule out the periodicities in $G$ claimed in ~\cite{Anderson}, by many orders of magnitude,
it is important to look for a variability in $G$ using independent terrestrial probes. 
This was the main motivation for D21 as well as this work.}

The main premise for the analysis in D21 is that the \dg measurements provide an upper bound to the fractional variation in $G$.
For this purpose, D21 analyzed the $\chi^2$ functional between the data and measurements, and looked for a minimum at 5.9 years for the sinusoidal functions. They do not find any such periodicity or  linear time-dependent variations. The limits which they obtain rule out the claims in ~\cite{Anderson}.

\begin{figure*}
    \centering
    \includegraphics[scale=0.5]{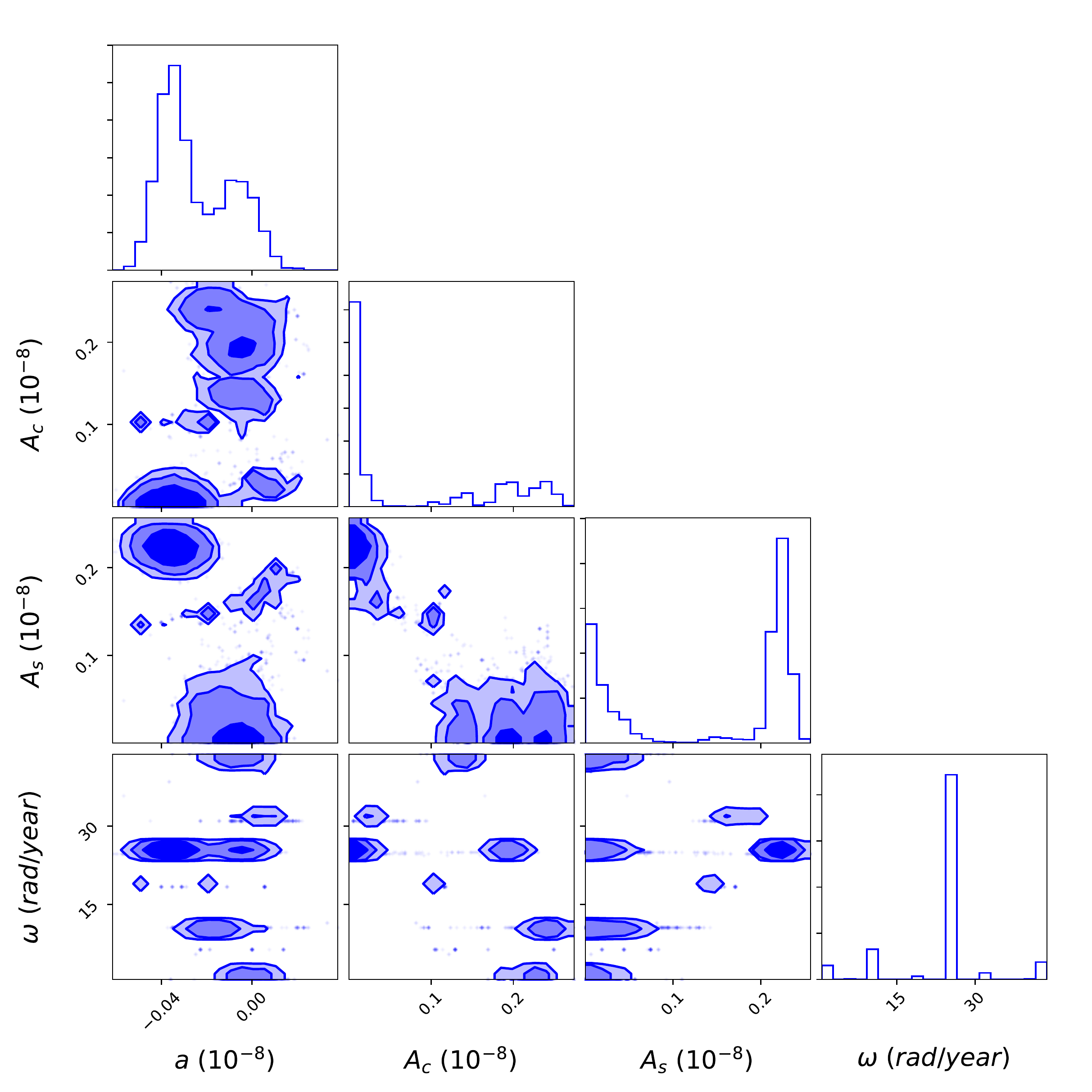}
    \caption{Contours showing 68\%, 90\%, and 95\% credible intervals for the free parameters specified in the  first sinusoidal function given in Equation \ref{eq:sinu1}}
    \label{fig:contour_sinu1}
\end{figure*}

\begin{figure*}
    \centering
    \includegraphics[scale=0.5]{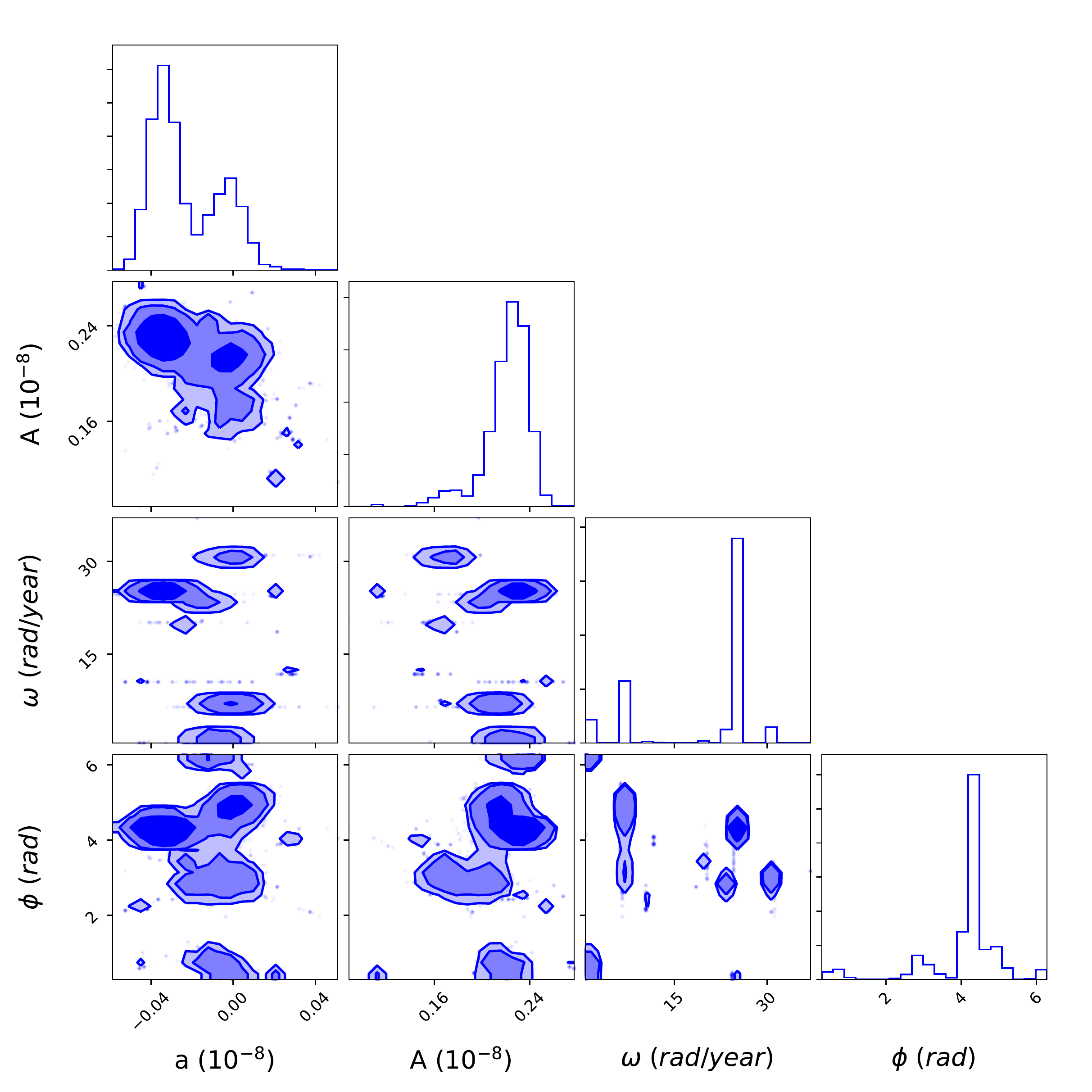}
    \caption{Contours showing the  68\%, 90\%, and 95\% credible intervals for the  free parameters in the  sinusoidal function given in Eq.~\ref{eq:sinu2}}
    \label{fig:contour_sinu2}
\end{figure*}

\begin{figure*}
    \centering
    \includegraphics[scale=0.5]{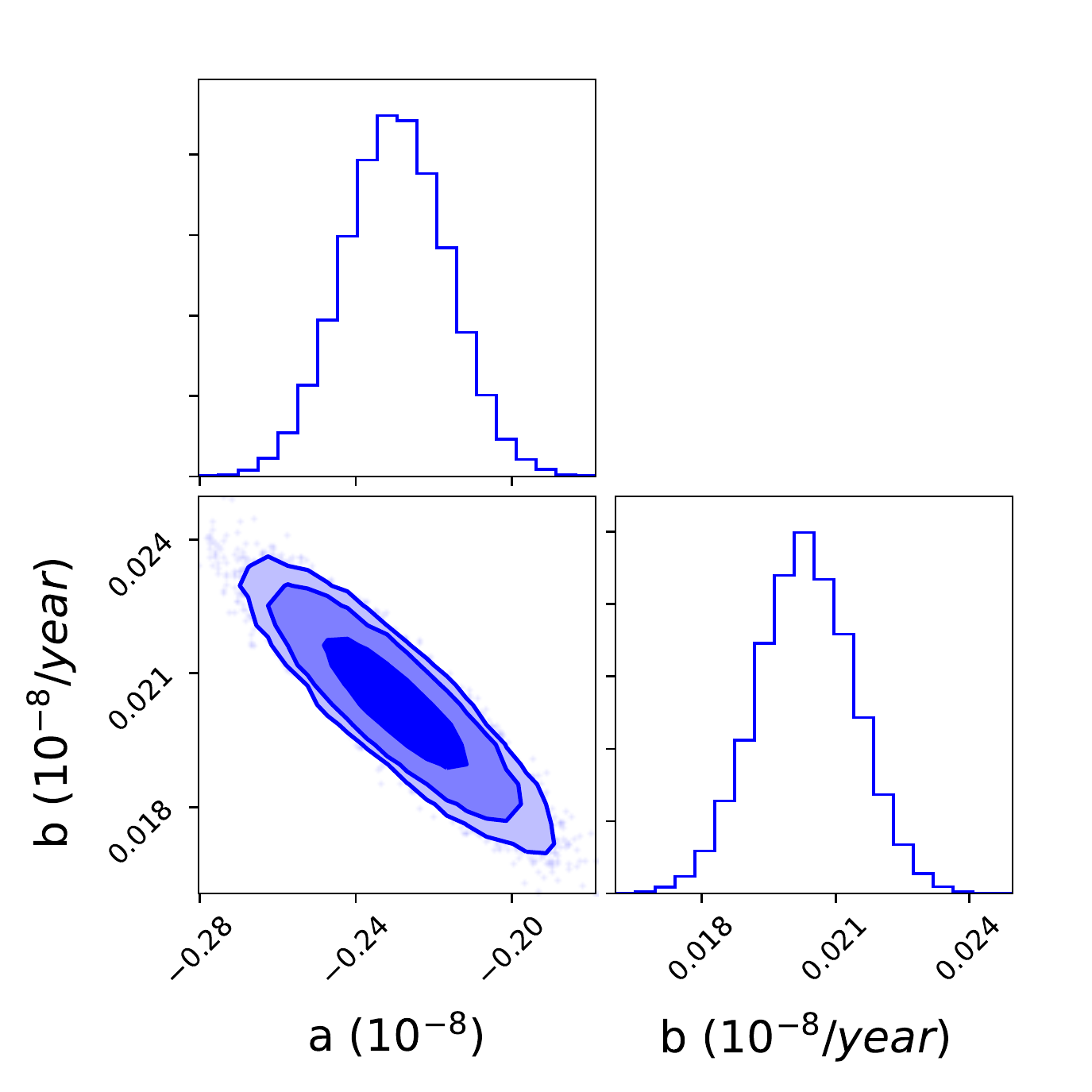}
    \caption{Contours containing the 68\%, 90\%, and 95\% credible intervals for the  free parameters in  the linear function specified in  Eq.~\ref{eq:linear}}
    \label{fig:contour_linear}
\end{figure*}

However,  D21 did not incorporate the errors in \dg, while evaluating the $\chi^2$. In this work, we improve upon the work in D21 by considering the measurement errors in all our analyses. We considered  four functions 
to model the putative behavior of \dg: a constant model, two sinusoidal models (Eq.~\ref{eq:sinu1} and Eq.~\ref{eq:sinu2}), and a linear model (Eq.~\ref{eq:linear}). For each of these models, we maximized a  log-likelihood (Eq.~\ref{eq:lnL}) function by incorporating the errors to get the allowed regions on the parameters. 
These allowed regions can be found in Fig.~\ref{fig:contour_sinu1}, Fig.~\ref{fig:contour_sinu2}, and Fig.~\ref{fig:contour_linear} for the two sinusoidal and linear models.
The best-fit values for $\chi^2$ for each of   these models is tabulated in Table~\ref{tab:my_label}.  As we see, the $\chi^2$/dof for all the models hitherto  analyzed in this work are greater than 20, indicating that none of them provide a good fit to the data.

Given these results, we therefore redo our search for a variability in \dg  by incorporating  an unknown intrinsic scatter, while fitting all the four models. The best-fit results  from these analyses are tabulated in Table~\ref{tab:scatter_res}.  The best-fit intrinsic scatter we obtain is about 2.8 times larger than the errors in  \dg measurements. We carry out a model comparison between the four models using AIC and BIC.
We find that the constant models are decisively favored over the sinusoidal ones. However, we find that the linear model is marginally favored over the constant model. \rthis{Additional data would be  needed to ascertain if the difference becomes significant.}

Based on our analysis without intrinsic scatter, we find from our fits to the two sinusoidal models that the relative variance of $G$ for any sinusoidal variation is $\leqslant 2 \times 10^{-9}$, which is about four orders of magnitude more stringent than the amplitude of periodic variations inferred from previous $G$ measurements.
Using the best-fit linear model, we can constrain $\frac{\dot{G}}{G} < 2.14 \times 10^{-10}$/year, which is about a factor of two more stringent than the corresponding limit obtained in D21. \rthis{Therefore, although our limits on variation of $G$ are not as stringent as those obtained from astrophysical observations, they are still slightly better than those in D21, after we incorporate the errors in measurements.}

To conclude, we concur with D21 that there is no evidence for a  periodicity or a linear variation in the \dg measurements, even after incorporating the errors, but without adding an intrinsic scatter. When we add an intrinsic scatter, the linear time evolution model for $G$ is marginally favored over a constant model.

\rthis{In the spirit of open science analysis, we have made all codes used in this work publicly available, which can be found at the github link \url{https://github.com/srinixbhagvati/Search-for-Variability-in-Newtons-Constant}}

\begin{table*}[h]
    \centering
    \begin{tabular}{|c|c|c|}
    \cline{1-3}
        Model &  $\omega$ Peaks  & $\chi^2/DOF$ values\\
         & ($rad/year$) & \\
    \cline{1-3}
        Constant & & $7786.6/354$\\
    \cline{1-3}
        First Sinusoidal & $\omega = 0.64$ ($T = 9.82$ $yrs$)& $7114.3/351$\\
        Model & $\omega = 10.56$ ($T = 0.59$ $yrs$) & $7195.5/351$\\
         & $\omega = 24.47$ ($T = 0.26$ $yrs$) & $7190.1/351$ \\
         & $\omega = 40.46$ ($T = 0.15$ $yrs$) & $7609.7/351$\\
    \cline{1-3}
        Second Sinusoidal & $\omega = 6.35$ ($T = 0.99$ $yrs$)& $7224.5/351$\\
        Model & $\omega = 24.50$ ($T = 0.26$ $yrs$) & $7174.3/351$ \\
    \cline{1-3}
        Linear Model & & $7443.1/353$ \\
    \cline{1-3}
        
    \end{tabular}
    \caption{The $\chi^2/DOF$ values for the best-fit parameters for the four models (Equations \ref{eq:const}, \ref{eq:sinu1}, \ref{eq:sinu2}, and \ref{eq:linear}) analyzed. The $\chi^2$ values are calculated for the sinusoidal models at the various peaks of $\omega$ from the contours shown in Figures \ref{fig:contour_sinu1} and \ref{fig:contour_sinu2}.  The best-fit values for the other parameters can be found in the text. As we can see, none of the four models provide an adequate fit to the observed \dg measurements.}
    \label{tab:my_label}
\end{table*}

\iffalse
\begin{figure*}
    \centering
    \includegraphics[scale=0.5]{contour4_scatter.pdf}
    \caption{Contours containing the 68\%, 90\%, and 95\% credible intervals for the  free parameters in  the constant function specified in  Eq.~\ref{eq:const} with added intrinsic scatter as described in Section \ref{sec:scatter}.}
    \label{fig:contour_const_scatter}
\end{figure*}

\begin{figure*}
    \centering
    \includegraphics[scale=0.5]{contour1_scatter.pdf}
    \caption{Contours containing the 68\%, 90\%, and 95\% credible intervals for the  free parameters in  the first sinusoidal function specified in  Eq.~\ref{eq:sinu1} with added intrinsic scatter as described in Section \ref{sec:scatter}.}
    \label{fig:contour_sinu1_scatter}
\end{figure*}

\begin{figure*}
    \centering
    \includegraphics[scale=0.5]{contour2_scatter.pdf}
    \caption{Contours containing the 68\%, 90\%, and 95\% credible intervals for the  free parameters in  the second sinusoidal function specified in  Eq.~\ref{eq:sinu2} with added intrinsic scatter as described in Section \ref{sec:scatter}.}
    \label{fig:contour_sinu2_scatter}
\end{figure*}

\begin{figure*}
    \centering
    \includegraphics[scale=0.5]{contour3_scatter.pdf}
    \caption{Contours containing the 68\%, 90\%, and 95\% credible intervals for the  free parameters in  the linear function specified in  Eq.~\ref{eq:linear} with added intrinsic scatter as described in Section \ref{sec:scatter}.}
    \label{fig:contour_linear_scatter}
\end{figure*}
\fi

\begin{table*}[t]
    \centering
    \begin{tabular}{|c|c|c|c|c|}
    \cline{1-5}
        Model &  $\omega$ Peaks  & $\chi^2/DOF$ values & $\Delta$AIC values & $\Delta$ BIC values\\
         & ($rad/year$) & & &\\
    \cline{1-5}
        Constant & & $354.4/353$ & & \\
    \cline{1-5}
        First Sinusoidal Model& $\omega = 6.10$ ($T = 1.03$ $yrs$)& $382.4/350$ & $34$ & $45$\\
        
    \cline{1-5}
        Second Sinusoidal Model& $\omega = 6.11$ ($T = 1.03$ $yrs$)& $375.7/350$ & $27$ & $39$\\
    \cline{1-5}
        Linear Model & & $348.1/352$ & $-4.3$ & $-0.4$ \\
    \cline{1-5}
        
    \end{tabular}
    \caption{The $\chi^2/DOF$ values for the best-fit parameters for the four models (Equations \ref{eq:const}, \ref{eq:sinu1}, \ref{eq:sinu2}, and \ref{eq:linear}) after incorporating an unknown intrinsic scatter $\sigma_{scatter}$.  We also calculate $\Delta$ AIC and $\Delta$ BIC between the constant and time-varying models. We find that the constant model is decisively favored over the sinusoidal  models. However, the linear model is marginally favored over the  constant one.}
    \label{tab:scatter_res}
\end{table*}

\begin{acknowledgments}
We are grateful to De-Chang Dai for systematically collating the  IGETS $\delta$g/g data and sharing it with us as well as for useful feedback on our manuscript. \rthis{We are also grateful to the anonymous referees for constructive feedback on this work.}
\end{acknowledgments}

\bibliography{paper.bib}

\end{document}